\documentclass[12pt]{amsart}
\usepackage{amssymb,latexsym}
\theoremstyle{plain}
\numberwithin{equation}{section}
\newtheorem{theorem}{Theorem}

\newtheorem{remark}{Remark}
\newtheorem{definition}{Definition}
\begin{document}

%archives090506.tex

\title[RHPWN and CFT]{\textbf{Renormalized Higher Powers of White Noise (RHPWN) and Conformal Field Theory}}

\author{Luigi Accardi}
\address{Centro Vito Volterra, Universit\`{a} di Roma Tor Vergata\\
            via Columbia  2, 00133 Roma, Italy}
\email{accardi@Volterra.mat.uniroma2.it}
\author{Andreas Boukas}
\address{Department of Mathematics and Natural Sciences, American College of Greece\\
 Aghia Paraskevi, Athens 15342, Greece}
\email{andreasboukas@acgmail.gr}

\date{}

\begin{abstract}   

The Virasoro--Zamolodchikov Lie algebra $w_{\infty}$ has been widely studied
in string theory and in conformal field theory, motivated by the attempts of developing a satisfactory theory of quantization of gravity. The renormalized higher powers of quantum white noise (RHPWN) $*$-Lie algebra has been recently investigated in quantum probability, motivated by the attempts to develop a nonlinear generalization of stochastic and white noise analysis. We prove that, after introducing a new renormalization technique, the RHPWN Lie algebra includes a second quantization of the $w_{\infty}$ algebra. Arguments discussed at the end of this note suggest the conjecture that this inclusion is in fact an identification

\end{abstract}

\maketitle

\section{Introduction}

We will use the notations of the paper \cite{1} which contains the proofs of all the results used in this section. 
The standard Boson white noise $*$--Lie algebra is defined by the commutation relations

\begin{eqnarray*}
[b_t,b_s^{\dagger}]&=&\delta(t-s)\cdot 1  \\
\lbrack b_t^{\dagger},b_s^{\dagger}\rbrack&=&[b_t,b_s]=0
\end{eqnarray*}
\begin{eqnarray*}
(b_s^{\dagger})^{\dagger} = b_s \qquad ; \qquad 1^{\dagger} = 1
\end{eqnarray*}
where $1$ (often omitted from the notations) denotes the central element and
all the identities are meant in the operator distribution sense described in \cite{0}.

The formal extension of the above commutation relations to the associative $*$--algebra 
generated by $b_t,b_s^{\dagger},1$ leads to the identities:

\begin{eqnarray}\label{1}
[{b_t^{\dagger}}^nb_t^k,{b_s^{\dagger}}^Nb_s^K]&=&
\epsilon_{k,0}\epsilon_{N,0}\sum_{L\geq 1} \binom{k}{L}N^{(L)}\,
{b_t^{\dagger}}^{n}\,{b_s^{\dagger}}^{N-L}\,b_t^{k-L}\,b_s^K\,
\delta^L(t-s)\\
&-&\epsilon_{K,0}\epsilon_{n,0}\sum_{L\geq 1} \binom{K}{L}n^{(L)}\,
{b_s^{\dagger}}^{N}\,{b_t^{\dagger}}^{n-L}\,b_s^{K-L}\,b_t^k\,
\delta^L(t-s)\nonumber
\end{eqnarray}
where $\forall n,k,N,K\in\mathbb{N}$

\begin{equation*}
\binom{K}{L} := {K!\over L!(K-L)! }\qquad \hbox{(binomial coefficient)} \ ; \ 
\binom{K}{L} = 0 \ \hbox{, if} \ K<L
\end{equation*}

\begin{eqnarray*}
 \epsilon_{n,k}:=1-\delta_{n,k}  \qquad \hbox{(Kronecker's delta)}
\end{eqnarray*}
\begin{equation*}
n^{(L)}:=n(n-1)\cdots (n-L+1)  \qquad ; \qquad n^{(0)}=1 \ ; \ n^{(L)}= 0 \ \hbox{, if} \ n<L
\end{equation*}

The right hand side of the above identity is ill defined because 
of the powers $\delta^L(t-s)$ of the $\delta$--function. 
Any procedure to give a meaning to these powers will be 
called a renormalization rule. In  the present note we will use the following renormalization rule whose motivations are discussed in \cite{3}:

\begin{eqnarray}\label{ff3}
\delta^l(t-s)=\delta(s)\,\delta(t-s),\,\,\,\,\,l=2,3, 4, \dots 
\end{eqnarray}
The right hand side of (\ref{ff3}) is well defined as a convolution of distributions. Using this (\ref{1}) can be rewritten in the form:
\begin{eqnarray}
&[{b_t^{\dagger}}^nb_t^k,{b_s^{\dagger}}^Nb_s^K]=&\label{5}\\
&\epsilon_{k,0}\epsilon_{N,0}\,\left( 
k\,N\, {b_t^{\dagger}}^{n}\,{b_s^{\dagger}}^{N-1}\,b_t^{k-1}\,b_s^K\,\delta(t-s) +    
\sum_{L\geq 2} \binom{k}{L}N^{(L)}\,{b_t^{\dagger}}^{n}
\,{b_s^{\dagger}}^{N-L}\,b_t^{k-L}\,b_s^K\, \delta(s)\,\delta(t-s) \right) &\nonumber\\
&-\epsilon_{K,0}\epsilon_{n,0}\left( 
K\,n \,{b_s^{\dagger}}^{N}\,{b_t^{\dagger}}^{n-1}\,b_s^{K-1}\,b_t^k\,\delta(t-s)   +   
\sum_{L\geq 2} \binom{K}{L}n^{(L)}\,
{b_s^{\dagger}}^{N}\,{b_t^{\dagger}}^{n-L}\,b_s^{K-L}\,b_t^k\,\delta(s)\,\delta(t-s)\right)&\nonumber
\end{eqnarray}

Introducing test functions and the associated smeared fields

\[
B_k^n(f):=\int_{\mathbb{R}^d}\,f(t)\,{b_t^{\dagger}}^n\,b_t^k\,dt
\]

the commutation relations (\ref{5}) become:

\begin{eqnarray}
&[B^n_k(g),B^N_K(f) ]=\left(\epsilon_{k,0}\epsilon_{N,0}\, 
k\,N- \epsilon_{K,0}\epsilon_{n,0}\, K\,n     \right)\, B^{N+n-1}_{K+k-1}(g f)& \label{6} \\
&+\sum_{L= 2}^{ (K\wedge n)\vee (k \wedge N)}\, \theta_L (n,k;N,K)\,
g(0) \,f(0)\,{b_0^{\dagger}}^{N+n-L}\,b_0^{K+k-L}   \nonumber
\end{eqnarray}

\begin{equation}
\theta_L(n,k;N,K):=\epsilon_{k, 0}\,\epsilon_{N, 0}\,\binom{k}{ L}\, N^{(L)} - \epsilon_{K, 0}\,\epsilon_{n, 0} \,\binom{K}{L} \, n^{(L)}\label{2a}
\end{equation}

which still contains the ill defined symbols ${b_0^{\dagger}}^{N+n-L}$ and $b_0^{K+k-L} $. However, if the test function space is chosen so that 
\begin{equation}\label{zeroatorig} 
f(0)=g(0)=0
\end{equation}
then the singular term in (\ref{6}) vanishes and the commutation relations (\ref{6}) become:
\begin{eqnarray}\label{8} 
[B^n_k(g),B^N_K(f) ]_{R}:= \left( k\,N- K\,n     \right)\, B^{n+N-1}_{k+K-1}(g f)
\end{eqnarray}
which no longer include ill defined objects. The symbol $[ \ \cdot \ , \ \cdot \ ]_{R}$
denotes the renormalized commutation relations. 

A simple direct calculation shows that the commutation relations (\ref{8}) define, 
on the family of symbols $B^n_k(f) $, a structure of $*$--Lie algebra with involution
$$
B^n_k(f)^* := B^k_n(\overline f)
$$

From the commutation relations (\ref{8}) it is clear that, fixing
a sub--set $I\subseteq \mathbb{R}^d$, not containing $0$, and the test function
\begin{equation}
\chi_I (s) =\begin{cases}   
1\ ,\quad s\in I\\
0\ ,\quad s\notin I
\end{cases}
\end{equation}
the commutation relations (\ref{8}) restricted to the (self--adjoint) family  
\begin{equation}\label{8a} 
\{ B^{n}_{k}:= B^{n}_{k}(\chi_I ) \ : \ n,k\in \mathbb{N} \, , \ n+k\geq 3   \}
\end{equation}
give
\begin{eqnarray}\label{8b} 
[B^n_k,B^N_K ]_{R}:= \left( k\,N- K\,n     \right)\, B^{n+N-1}_{k+K-1}
\end{eqnarray}
The arguments in (\cite{3}) then suggest the natural interpretation of the $*$--Lie--algebra, defined by the relations (\ref{8a}), (\ref{8b}), as the {\it $1$--mode algebra} of the RHPWN and, conversely, of the  
RHPWN $*$--Lie--algebra as a current algebra of its $1$--mode version.

Now recall the following definition (see \cite{5}-\cite{7}):

\begin{definition} 
The $w_{\infty}-*$--Lie--algebra is the infinite dimensional Lie 
algebra spanned by the generators $\hat B^n_k$, where $n\geq2$ 
and $k\in\mathbb{Z}$, with commutation relations:
\begin{equation}\label{extvir} 
[ \hat B^n_k , \hat B^N_K ]_{w_{\infty} } =\left(   (N-1)\,k-(n-1)\,K  \right) \,\hat B^{n+N-2}_{k+K}
\end{equation}
and involution
\begin{equation} \label{adjvir} 
\left( \hat{B}^n_k \right)^*= \hat{B}^n_{-k}
\end{equation}
\end{definition}

\begin{remark}
The $w_{\infty}-*$--Lie--algebra, whose elements are interpreted 
as area preserving diffeomorphisms of $2$--manifolds, contains 
as a sub--Lie--algebra (not as $*$--algebra) the (centerless) 
Virasoro (or Witt) algebra with commutations relations
\[
[\hat{B}^2_k(g),\hat{B}^2_{K}(f) ]_{V}:=(k-K)\,\hat{B}^{2}_{k+k}(g f)
\] 
Both $w_{\infty}$ and a quantum deformation of it, denoted 
$W_{\infty}$ and defined as a (non-unique) large  $N$ limit of Zamolodchikov's $W_N$ algebra (\cite{8}), have been studied extensively  
(\cite{5}, \cite{6}, \cite{ketov}, \cite{4b}, \cite{7}) in connection to two-dimensional Conformal Field Theory and Quantum Gravity. 
\end{remark}

The striking similarity between the commutation relations (\ref{extvir})
and (\ref{8b}) suggests that the two algebras are deeply related. The following theorem shows that the  current algebra, over $\mathbb{R}^d$, 
of the $w_{\infty}-*$--Lie--algebra can be realized in terms of the renormalized powers of white noise. The converse of this statement is intuitively obvious at the level of formal white noise operators, but
a precise statement of this last statement will be discussed elsewhere.

\begin{theorem}\label{comw} 
Let $\mathcal{S}_0$ be the test function space of complex valued (right-continuous) step functions on $\mathbb{R}^d$ assuming a finite number of values and vanishing at zero, and let the powers of the delta function be renormalized by the prescription 
\begin{eqnarray}\label{f3}
\delta^l(t-s)=\delta(s)\,\delta(t-s),\,\,\,\,\,l=2,3,... 
\end{eqnarray}
(cf. \cite{3}). Then the white noise operators
\begin{equation}\label{op}
\hat{B}_k^n(f):=
\int_{\mathbb{R}^d}\,f(t)\,e^{ \frac{k}{2}(b_t- b_t^{\dagger})}
\left(\frac{ b_t+ b_t^{\dagger}}{2}\right)^{n-1} \,  
e^{ \frac{k}{2}(b_t- b_t^{\dagger})}\,dt
\end{equation}
satisfy the relations  (\ref{extvir}) and (\ref{adjvir}) of the 
$w_{\infty}$--Lie algebra.
\end{theorem}
\begin{remark}
The integral on the right hand side of (\ref{op}) is meant in the 
sense that one expands the exponential series, applies the commutation relations (\ref{1}) to bring the resulting expression to normal order,
introduces the renormalization prescription (\ref{f3}), integrates the resulting expressions after multiplication by a test function and interprets the result  as a quadratic form on the exponential vectors.
\end{remark}
\begin{proof} 
The relation (\ref{adjvir}) is obvious, thus we will only prove 
(\ref{extvir}). To this goal notice that the left hand side 
of (\ref{extvir}) is equal to:
\[
\int_{\mathbb{R}^d}\int_{\mathbb{R}^d}g(t)f(s)\left[
e^{ \frac{k}{2}(b_t- b_t^{\dagger})}
\left(\frac{ b_t+ b_t^{\dagger}}{2}\right)^{n-1}   
e^{ \frac{k}{2}(b_t- b_t^{\dagger})}, 
e^{ \frac{K}{2}(b_s- b_s^{\dagger})}
\left(\frac{ b_s+ b_s^{\dagger}}{2}\right)^{N-1}   
e^{ \frac{K}{2}(b_s- b_s^{\dagger})}\right] \,dt\,ds
\]
\[
=\int_{\mathbb{R}^d}\int_{\mathbb{R}^d}g(t)f(s)\, e^{ \frac{k}{2}(b_t- b_t^{\dagger})}\left(\frac{ b_t+ b_t^{\dagger}}{2}\right)^{n-1}   e^{ \frac{k}{2}(b_t- b_t^{\dagger})}  e^{ \frac{K}{2}(b_s- b_s^{\dagger})}\left(\frac{ b_s+ b_s^{\dagger}}{2}\right)^{N-1}   e^{ \frac{K}{2}(b_s- b_s^{\dagger})}    \,dt\,ds
\]
\[
-\int_{\mathbb{R}^d}\int_{\mathbb{R}^d}g(t)f(s)\, e^{ \frac{K}{2}(b_s- b_s^{\dagger})}\left(\frac{ b_s+ b_s^{\dagger}}{2}\right)^{N-1}   e^{ \frac{K}{2}(b_s- b_s^{\dagger})}  e^{ \frac{k}{2}(b_t- b_t^{\dagger})}\left(\frac{ b_t+ b_t^{\dagger}}{2}\right)^{n-1}   e^{ \frac{k}{2}(b_t- b_t^{\dagger})}   \,dt\,ds
\]
Since $[b_t- b_t^{\dagger} ,b_s - b_s^{\dagger} ]=0$, this is equal to:
\[
=\int_{\mathbb{R}^d}\int_{\mathbb{R}^d}g(t)f(s)\, e^{ \frac{k}{2}(b_t- b_t^{\dagger})}\left(\frac{ b_t+ b_t^{\dagger}}{2}\right)^{n-1}    e^{ \frac{K}{2}(b_s- b_s^{\dagger})} e^{ \frac{k}{2}(b_t- b_t^{\dagger})}\left(\frac{ b_s+ b_s^{\dagger}}{2}\right)^{N-1}   e^{ \frac{K}{2}(b_s- b_s^{\dagger})}    \,dt\,ds
\]

\[
-\int_{\mathbb{R}^d}\int_{\mathbb{R}^d}g(t)f(s)\, e^{ \frac{K}{2}(b_s- b_s^{\dagger})}\left(\frac{ b_s+ b_s^{\dagger}}{2}\right)^{N-1}    e^{ \frac{k}{2}(b_t- b_t^{\dagger})}e^{ \frac{K}{2}(b_s- b_s^{\dagger})} \left(\frac{ b_t+ b_t^{\dagger}}{2}\right)^{n-1}   e^{ \frac{k}{2}(b_t- b_t^{\dagger})}   \,dt\,ds
\]

\[
=\frac{ 1  }{ 2^{n+N-2}}\{\int_{\mathbb{R}^d}\int_{\mathbb{R}^d}g(t)f(s)\, e^{ \frac{k}{2}(b_t- b_t^{\dagger})}(b_t+ b_t^{\dagger})^{n-1}    e^{ \frac{K}{2}(b_s- b_s^{\dagger})} e^{ \frac{k}{2}(b_t- b_t^{\dagger})}( b_s+ b_s^{\dagger})^{N-1}   e^{ \frac{K}{2}(b_s- b_s^{\dagger})}    \,dt\,ds
\]

\[
-\int_{\mathbb{R}^d}\int_{\mathbb{R}^d}g(t)f(s)\, e^{ \frac{K}{2}(b_s- b_s^{\dagger})}( b_s+ b_s^{\dagger})^{N-1}    e^{ \frac{k}{2}(b_t- b_t^{\dagger})}e^{ \frac{K}{2}(b_s- b_s^{\dagger})} ( b_t+ b_t^{\dagger})^{n-1}   e^{ \frac{k}{2}(b_t- b_t^{\dagger})}   \,dt\,ds \}
\]
From the formal expression (\ref{1}) of the CCR we deduce the identities:
\[
e^{\frac{K}{2}(b_s- b_s^{\dagger})}(b_t+ b_t^{\dagger})^{n-1}  =\sum_{m=0}^{n-1}\binom{n-1}{m}(b_t+ b_t^{\dagger})^{m}K^{n-1-m}\delta^{n-1-m}(t-s)\,e^{\frac{K}{2}(b_s- b_s^{\dagger})}
\]

\[
  e^{\frac{k}{2}(b_t- b_t^{\dagger})}(b_s+ b_s^{\dagger})^{N-1}=\sum_{m=0}^{N-1}\binom{N-1}{m}(b_s+ b_s^{\dagger})^{m}k^{N-1-m}\delta^{N-1-m}(t-s)\,e^{\frac{k}{2}(b_t- b_t^{\dagger})}
\]

\[
(b_t+ b_t^{\dagger})^{n-1}e^{\frac{K}{2}(b_s- b_s^{\dagger})}  =e^{\frac{K}{2}(b_s- b_s^{\dagger})}\,\sum_{m=0}^{n-1}\binom{n-1}{m}(b_t+ b_t^{\dagger})^{m}(-1)^{n-1-m}K^{n-1-m}\delta^{n-1-m}(t-s)
\]

\[
 (b_s+ b_s^{\dagger})^{N-1}e^{\frac{k}{2}(b_t- b_t^{\dagger})}  =e^{\frac{k}{2}(b_t- b_t^{\dagger})}\, \sum_{m=0}^{N-1}\binom{N-1}{m}(b_s+ b_s^{\dagger})^{m}(-1)^{N-1-m}k^{N-1-m}\delta^{N-1-m}(t-s)
\]
These identities imply that:
\[
[\hat{B}^n_k(g),\hat{B}^N_K(f) ]=
\frac{ 1  }{ 2^{n+N-2}}\{\sum_{m_1=0}^{n-1}\sum_{m_2=0}^{N-1}
\binom{n-1}{m_1}\binom{N-1}{m_2}(-1)^{n-1-m_1}K^{n-1-m_1}k^{N-1-m_2}
\]
\[
\times\int_{\mathbb{R}^d}\int_{\mathbb{R}^d}g(t)f(s)\,
e^{\frac{k}{2}(b_t- b_t^{\dagger})}e^{\frac{K}{2}(b_s- b_s^{\dagger})}
(b_t+ b_t^{\dagger})^{m_1} (b_s+ b_s^{\dagger})^{m_2} 
e^{\frac{k}{2}(b_t- b_t^{\dagger})}e^{\frac{K}{2}(b_s- b_s^{\dagger})} 
\]
\[
\times \delta^{n-1-m_1 + N-1-m_2}(t-s)\,dt\,ds
\]

\[
-\sum_{m_3=0}^{N-1}\sum_{m_4=0}^{n-1}
\binom{N-1}{m_3}\binom{n-1}{m_4}(-1)^{N-1-m_3}k^{N-1-m_3}K^{n-1-m_4}
\]
\[
\times\int_{\mathbb{R}^d}\int_{\mathbb{R}^d}g(t)f(s)\,
e^{\frac{K}{2}(b_s- b_s^{\dagger})}e^{\frac{k}{2}(b_t- b_t^{\dagger})}
(b_s+ b_s^{\dagger})^{m_3} (b_t+ b_t^{\dagger})^{m_4} 
e^{\frac{K}{2}(b_s- b_s^{\dagger})}e^{\frac{k}{2}(b_t- b_t^{\dagger})}
\]
\[
\times \delta^{N-1-m_3 + n-1-m_4}(t-s)\,dt\,ds \}
\]

The term ($m_1=n-1$, $m_2=N-1$) cancels out with ($m_3=N-1$, $m_4=n-1$).  

The renormalization prescription (\ref{f3}) and the choice of test functions vanishing at zero imply that
$$
\sum_{m_1=0}^{n-3}\sum_{m_2=0}^{N-3} (\dots )=
 \sum_{m_3=0}^{N-3}\sum_{m_4=0}^{n-3} (\dots ) = 0
$$ 
Therefore, after the renormalization prescription (\ref{f3}),  
the only surviving terms are those corresponding to the pairs
$$
(m_1=n-1, m_2=N-2) \quad , \quad (m_3=N-1, m_4=n-2)
$$
$$
(m_1=n-2, m_2=N-1)  \quad , \quad (m_3=N-2 , m_4=n-1) 
$$
and we obtain:
\[
[\hat{B}^n_k(g),\hat{B}^N_K(f) ]=
\]
\[
=\frac{1}{ 2^{n+N-2}}((N-1)k-(n-1)K-(n-1)K+(N-1)k)
\]
\[
\times \int_{\mathbb{R}^d}\,g(t)f(t)\,e^{ \frac{k+K}{2}(b_t- b_t^{\dagger})}( b_t+ b_t^{\dagger})^{n+N-3}  e^{ \frac{k+K}{2}(b_t- b_t^{\dagger})}\,dt
\]
\[
=\frac{2}{ 2^{n+N-2}}((N-1)k-(n-1)K)\int_{\mathbb{R}^d}\,g(t)f(t)\,e^{ \frac{k+K}{2}(b_t- b_t^{\dagger})}( b_t+ b_t^{\dagger})^{n+N-3}    e^{ \frac{k+K}{2}(b_t- b_t^{\dagger})}\,dt
\]
\[
=\frac{1}{ 2^{n+N-3}}((N-1)k-(n-1)K)\int_{\mathbb{R}^d}\,g(t)f(t)\,e^{ \frac{k+K}{2}(b_t- b_t^{\dagger})}( b_t+ b_t^{\dagger})^{n+N-3}    e^{ \frac{k+K}{2}(b_t- b_t^{\dagger})}\,dt
\]
\[
=(k(N-1)-K(n-1))\hat{B}^{n+N-2}_{k+K}(g f)
\]

\end{proof}

\end{document}